\documentclass[aps,prl,superscriptaddress,twocolumn]{revtex4-1}
\usepackage[utf8]{inputenc}
\usepackage{dsfont}
\usepackage{amsfonts}
\usepackage{amsmath}
\allowdisplaybreaks[4]        
\usepackage{amssymb}
\usepackage{euscript}     
\usepackage{braket}
\usepackage{starfont}
\usepackage{color,soul}         
\usepackage{tensor}        
\usepackage{amsthm}
\usepackage{graphicx}
\usepackage{slashed}
\usepackage{leftidx}
\usepackage{subfigure}
\usepackage{bbm}
\definecolor{outerspace}{rgb}{0.25, 0.29, 0.3}
\definecolor{scarlet}{rgb}{1.0, 0.13, 0.0}
\usepackage[header,title,page,titletoc]{appendix}  
\definecolor{princetonorange}{rgb}{1.0, 0.56, 0.0}
\definecolor{WildStrawberry}{rgb}{1.0, 0.26, 0.64}
\definecolor{rossocorsa}{rgb}{0.83, 0.0, 0.0}
\definecolor{navyblue}{rgb}{0.0, 0.0, 0.5}
\usepackage{float}
\usepackage[paper=letterpaper,margin=1in]{geometry}
\usepackage{mathtools}

\DeclareMathAlphabet{\pazocal}{OMS}{zplm}{m}{n}
\newcommand{\req}[1]{(\ref{#1})} %{Eq.\thinspace(\ref{#1})}  

\newcommand{\bea}{\begin{eqnarray}}

\newcommand{\eea}{\end{eqnarray}}
\newcommand{\ba}{\begin{eqnarray}}
\newcommand{\ea}{\end{eqnarray}}

\newcommand{\be}{\begin{equation}}
\newcommand{\ee}{\end{equation} }
\newcommand{\beqa}{\begin{eqnarray}}
\newcommand{\eeqa}{\end{eqnarray}}
\newcommand{\beqar}{\begin{eqnarray*}}
\newcommand{\eeqar}{\end{eqnarray*}}

\renewcommand{\req}[1]{(\ref{#1})}

\makeatletter
\newcommand{\dal}{\mathop{\mathpalette\dal@\relax}}
\newcommand{\dal@}[2]{%
  \begingroup
  \sbox\z@{$\m@th#1\square$}%
  \dimen0=\fontdimen8
    \ifx#1\displaystyle\textfont\else
    \ifx#1\textstyle\textfont\else
    \ifx#1\scriptstyle\scriptfont\else
    \scriptscriptfont\fi\fi\fi3
  \makebox[\wd\z@]{%
    \hbox to \ht\z@{%
      \vrule width \dimen0
      \kern-\dimen0
      \vbox to \ht\z@{
        \hrule height \dimen0 width \ht\z@
        \vss
        \hrule height 2\dimen0
      }%
      \kern-2.5\dimen0
      \vrule width 2.5\dimen0
    }%
  }%
  \endgroup
}
\makeatother

\usepackage{hyperref}
\hypersetup{
    colorlinks,
    citecolor=blue,
    filecolor=navyblue,
    linkcolor=navyblue,
    urlcolor=navyblue,
    breaklinks=true
}
\begin{document}

\title{Exact electromagnetic duality with nonminimal couplings}
\author{Pablo A. Cano}
\email{pabloantonio.cano@kuleuven.be}
\affiliation{Instituut voor Theoretische Fysica, KU Leuven. Celestijnenlaan 200D, B-3001 Leuven, Belgium. }

\author{\'Angel Murcia}
\email{angel.murcia@csic.es}
\affiliation{Instituto de F\'isica Te\'orica UAM/CSIC. C/ Nicol\'as Cabrera 13-15, C.U. Cantoblanco, E-28049 Madrid, Spain}
\affiliation{Department of Mathematics, University of Hamburg. Bundesstra{\ss}e 55, D-20146 Hamburg, Germany}

\date{May 20, 2021}

\begin{abstract}

We study nonminimal extensions of Einstein-Maxwell theory with exact electromagnetic duality invariance. Any such theory involves an infinite tower of higher-derivative terms whose computation and summation usually represents a challenging problem. Despite that,  we manage to obtain a closed form of the action for all the theories with a quadratic dependence on the vector field strength. In these theories we find that the Maxwell field couples to gravity through a curvature-dependent susceptibility tensor that takes a peculiar form, reminiscent of that of Born-Infeld Lagrangians.
We study the static and spherically symmetric black hole solutions of the simplest of these models, showing that the corresponding equations of motion are invariant under rotations of the electric and magnetic charges. We  compute the perturbative corrections to the Reissner-Nordstr\"om solution in this theory, and in the case of extremal black holes we determine exactly the near-horizon geometry as well as the entropy. Remarkably, the entropy only possesses a constant correction despite the action containing an infinite number of terms. In addition, we find there is a lower bound for the charge and the mass of extremal black holes. When the sign of the coupling is such that the weak gravity conjecture is satisfied, the area and the entropy of extremal black holes vanish at the minimal charge. 

%carry out this summation and

%\tcb{We study nonminimal extensions of Einstein-Maxwell theory with exact electromagnetic duality invariance. Any such theory involves an infinite tower of higher-derivative terms whose computation usually represents an inaccessible problem, but we manage to obtain a closed form of the action for theories with quadratic dependence on the vector field strength. In that case, we show that the Maxwell field couples to gravity through a curvature-dependent susceptibility tensor that has a particular form. }
%We study the static and spherically symmetric black hole solutions of the simplest of these models, showing that they are invariant under rotations of the electric and magnetic charges. We explicilty compute the perturbative corrections to the Reissner-Nordstr\"om solution in this theory, and in the case of extremal black holes we determine exactly the near-horizon geometry as well as the entropy. Remarkably, the entropy only possesses a constant correction despite the action containing an infinite number of terms. In addition, we find there is a lower bound for the charge and the mass of extremal black holes. When the sign of the coupling is such that the weak gravity conjecture is satisfied, the area and the entropy of extremal black holes vanish at the minimal charge. 

\end{abstract}
\maketitle

Symmetries are a powerful guide to constrain the possible corrections to low-energy effective actions. In the case of Einstein-Maxwell (EM) theory, those symmetries include diffeomorphism and gauge invariance, but the vacuum equations of motion are additionally invariant under $\mathrm{SO}(2)$ duality rotations. Electromagnetic duality also plays a prominent role in supergravity and string theory \cite{Henneaux:1988gg,Duff:1989tf,Duff:1990hn,Sen:1992fr,Schwarz:1993vs,Ortin:2015hya}, and it is expected to be preserved by the higher-derivative terms that arise in these theories \cite{Sen:1992fr,deWit:2001pz,Chemissany:2006qd}. 
Therefore, seeking for modifications of EM theory that respect duality is more than justified.

One aspect of electromagnetic duality that distinguishes it from usual symmetries and that makes it particularly interesting is the fact that it is nonlinear. This means that, unlike in the case of linearly realized symmetries, one cannot simply find a basis of duality-invariant operators and add them to the action weighted by a set of arbitrary couplings. Instead, any duality-invariant deformation of EM theory implies an infinite series of terms.

To illustrate this, consider EM theory extended with four-derivative terms,
\begin{equation}
S=\frac{1}{16\pi}\int d^4x\sqrt{|g|}\left\{R-F_{\mu\nu}F^{\mu\nu}+\alpha\Delta\mathcal{L}\right\}\, ,
\label{eq:EM}
\end{equation}
where $R$ is the Ricci scalar of the metric $g_{\mu\nu}$, $F_{\mu\nu}=2\partial_{[\mu}A_{\nu]}$ is the field strength of the vector $A_{\mu}$, and the correction $\Delta\mathcal{L}$ is controlled by a constant $\alpha$ with dimensions of length$^2$. As shown in Ref.~\cite{Cano:2021tfs}, there are two four-derivative Lagrangians involving field strengths that are consistent with $\mathrm{SO}(2)$ duality, namely, 
\begin{equation}
\Delta\mathcal{L}_1=T_{\mu\nu}T^{\mu\nu}\, ,\quad \Delta\mathcal{L}_2=R_{\mu\nu}T^{\mu\nu}\, ,
\end{equation}
where $T_{\mu\nu}=F_{\mu \alpha} F_{\nu}{}^\alpha-\frac{1}{4}g_{\mu \nu} F^2$ is the Maxwell stress-energy tensor. However, such extensions only preserve duality to leading order in $\alpha$; in order to restore duality as an exact symmetry one needs to include an infinite tower of additional higher-derivative terms. 
 In a way, we can say that duality dictates how the theory should be completed, although such completion is not unique, since at each order one can introduce new independent terms that respect duality. 
In the case of the term $T_{\mu\nu}T^{\mu\nu}$, the simplest duality-invariant completion corresponds to (Einstein-)Born-Infeld theory \cite{Born:1934gh,Schrodinger:1935oqa,Gibbons:2001gy}, a model that has been known for a long time and which is a paradigmatic example of a nonlinear duality invariant theory --- see also \cite{BialynickiBirula:1992qj,Gibbons:1995cv,Gibbons:1995ap,Gaillard:1997zr,Gaillard:1997rt,Hatsuda:1999ys,Brace:1999zi,Chemissany:2011yv,BabaeiVelni:2016qea,Bandos:2020jsw} for other nonlinear models. 
On the other hand, to the best of our knowledge, exactly invariant theories with nonminimal couplings to gravity have not been studied so far.  Nonminimal couplings are nevertheless interesting as they generically appear in stringy effective actions --- see \textit{e.g.} \cite{Elgood:2020xwu} for an explicit example. Thus, the goal of this paper is to fill this gap, and we do so by studying the simplest duality-invariant completion of the Lagrangian $R_{\mu\nu}T^{\mu\nu}$. \\

\textbf{Quadratic theories.}
Since the correction $R_{\mu\nu}T^{\mu\nu}$ is quadratic in the field strength, we can find completions that respect this property, and hence we stick to theories of the form 

\begin{equation}
S=\frac{1}{16\pi}\int d^4x\sqrt{|g|}\left\{R-\chi^{\mu\nu\rho\sigma}F_{\mu\nu}F_{\rho\sigma}\right\}\, ,
\label{eq:theory}
\end{equation}
where $\chi^{\mu\nu\rho\sigma}$ is the susceptibility tensor, which is built out of the metric and the curvature.  It was shown in Ref.~\cite{Cano:2021tfs} that, in order for this theory to preserve duality, the susceptibility tensor must satisfy the identity $\tensor{(\star\chi)}{_{\mu\nu}^{\alpha\beta}}\tensor{(\star\chi)}{_{\alpha\beta}^{\rho\sigma}}=-\tensor{\delta}{_{[\mu}^{[\rho}}\tensor{\delta}{_{\nu]}^{\sigma]}}$, where $\tensor{(\star\chi)}{_{\alpha\beta}^{\rho\sigma}}=\tfrac{1}{2}\epsilon_{\alpha\beta\mu\nu}\tensor{\chi}{^{\mu\nu\rho\sigma}}$. To find tensors with this property, it proves useful to expand $\chi$ in powers of $\alpha$,

\begin{equation}
\tensor{\chi}{_{\mu\nu}^{\rho\sigma}}=\tensor{\delta}{_{[\mu}^{[\rho}}\tensor{\delta}{_{\nu]}^{\sigma]}}+\sum_{n=1}^{\infty}\alpha^{n}\tensor{\chi}{^{(n)}_{\mu\nu}^{\rho\sigma}}\, .
\end{equation}
so that one recovers Einstein-Maxwell theory when $\alpha\rightarrow 0$, and where each term $\tensor{\chi}{^{(n)}_{\mu\nu}^{\rho\sigma}}$ contains $n$ powers of the curvature.  The four-derivative term should match $\Delta\mathcal{L}_2$, and therefore we must choose 
\begin{align}
\tensor{\chi}{^{(1)}_{\mu\nu}^{\rho\sigma}}&=-\tensor{\hat{R}}{_{[\mu}^{[\rho}}\tensor{\delta}{_{\nu]}^{\sigma]}}\, ,
\label{eq:chi1}
\end{align}
where $\tensor{\hat{R}}{_{\mu}^{\rho}}=\tensor{R}{_{\mu}^{\rho}}-\tfrac{1}{4}R\tensor{\delta}{_{\mu}^{\rho}}$ is the traceless part of the Ricci tensor. Then, the simplest completion to an exactly duality-invariant theory can be obtained by choosing the rest of the terms as   
\begin{equation}
\tensor{\chi}{^{(n+1)}_{\mu\nu}^{\rho\sigma}}=\frac{1}{2}\sum_{p=1}^{n}\tensor{\left( \star \chi^{(p)} \right)}{_{\mu \nu}^{\alpha\beta}}\tensor{ \left( \star \chi^{(n+1-p)} \right)}{_{\alpha \beta}^{\rho\sigma}}\, .
\label{recursive1}
\end{equation}
This result is obtained from Eq.~(4.31) of \cite{Cano:2021tfs} after noticing the following identity, valid for tensors $Q^{(1)}_{\mu \nu \rho \sigma}$ and $Q^{(2)}_{\mu \nu \rho \sigma}$ which are antisymmetric in the indices $\{\mu \nu\}$ and $\{\rho \sigma\}$ but symmetric in the exchange of these pairs of indices\footnote{One must ensure that the different $\tensor{\chi}{^{(n)}_{\mu\nu}^{\rho\sigma}}$ are symmetric in the exchange of the pairs of indices $\{\mu \nu\}$ and $\{\rho \sigma\}$. This is guaranteed by the construction of the complete $\tensor{\chi}{_{\mu \nu}^{\rho \sigma}}$, but if one wants not to rely on separate results, it is possible to prove \eqref{recursive1} and that the different $\tensor{\chi}{^{(n)}_{\mu\nu}^{\rho\sigma}}$ have the appropriate symmetries through a simple induction process.}:
\begin{equation}
\tensor{(\star Q)}{^{(1)}_{\mu \nu}^{\alpha\beta}}\tensor{ (\star Q)}{^{(2)}_{\alpha \beta}^{\rho\sigma}}=-6\tensor{ Q}{^{(1)}_{[\alpha \beta}^{\alpha\beta}}\tensor{ Q}{^{(2)}_{\mu \nu]}^{\rho\sigma}}\,.
\end{equation}
The recursive relations \eqref{recursive1} allow us to write the Lagrangian at arbitrary orders in the curvature. In fact, all the previous terms $\tensor{\chi}{^{(n)}_{\mu\nu}^{\rho\sigma}}$ can be explicitly summed to yield a fully-non perturbative duality-invariant theory. For that we first note the following results, obtained through direct computation:
\begin{align}
\label{eq:dualprop}
\epsilon_{\mu \nu \alpha \beta} \tensor{\chi}{^{(1)}_{}^{\alpha \beta}_{\rho \sigma}}&=- \tensor{\chi}{^{(1)}_{\mu\nu}^{\alpha \beta}}\epsilon_{\alpha\beta \rho \sigma}\, ,\\
\label{eq:chi2facil}
\tensor{\chi}{^{(2)}_{\mu\nu}^{\rho\sigma}}&=\frac{1}{2} \tensor{\chi}{^{(1)}_{\mu\nu}^{\alpha \beta}} \tensor{\chi}{^{(1)}_{\alpha \beta}^{\rho\sigma}}\,.
\end{align}
Now let $b_{n}$ denote the $n$-th coefficient of the Taylor series $\sqrt{1+x^2}=\sum_{n=0}^\infty  b_n x^{2n}$. We have that $b_0=1$, $b_1=1/2$ and
\begin{equation}
b_n=(-1)^{n+1}\frac{(2n-3)!}{n!(n-2)! 2^{2n-2}}\, , \quad n>1\,.
\end{equation}
\noindent
We note that these coefficients satisfy the property
\begin{equation}\label{recbn}
b_{n+1}= -1/2\sum_{p=1}^{n} b_p b_{n+1-p}\, ,
\end{equation}
which we will need later. 
\noindent
We are going to prove that that for $n >1$:
\begin{align}
\label{eq:eventerms}
\tensor{\chi}{^{(2n)}_{\mu\nu}^{\rho\sigma}}&=b_n \left(  \tensor{\chi}{^{(1)}}\right)^{2n}{}_{\mu \nu}{}^{\rho \sigma}\,,\\ 
\tensor{\chi}{^{(2n+1)}_{\mu\nu}^{\rho\sigma}}&=0\,,
\label{eq:oddterms}
\end{align}
where we have defined
\begin{align}
\nonumber
 \left(  \tensor{\chi}{^{(1)}}\right)^{k}{}_{\mu \nu}{}^{\rho \sigma}&=\tensor{\chi}{^{(1)}_{\mu \nu}^{\alpha_1 \beta_1}}  \tensor{\chi}{^{(1)}_{\alpha_1 \beta_1 }^{\alpha_2 \beta_2}} \cdots \\& \cdots \tensor{\chi}{^{(1)}_{\alpha_{k-1} \beta_{k-1}}^{\rho \sigma}}\, , \quad k>1\,.
\end{align}
The proof for \eqref{eq:eventerms} can be done by induction. First, we notice that \eqref{eq:chi2facil} guarantees that \eqref{eq:eventerms} is true for $n=1$. Next assume that it is valid for generic $n$. For $m, p \in 2\mathbb{N}$ such that $m+p =2n+2$ we find that
\begin{widetext}
\begin{align}
\nonumber
\tensor{\left( \star \chi^{(p)} \right)}{_{\mu \nu}^{\alpha\beta}}\tensor{ \left( \star \chi^{(m)} \right)}{_{\alpha \beta}^{\rho\sigma}}&= b_p b_m \tensor{\left( \star \chi^{(1)} \right)}{_{\mu \nu}^{\lambda \gamma }}  \left(  \tensor{\chi}{^{(1)}}\right)^{p-1}{}_{\lambda \gamma}{}^{\alpha \beta }  \tensor{\left( \star \chi^{(1)} \right)}{_{\alpha \beta}^{\eta \kappa}} \left(  \tensor{\chi}{^{(1)}}\right)^{m-1}{}_{\eta \kappa}{}^{\rho \sigma }\\&=- b_p b_m \left(  \tensor{\chi}{^{(1)}}\right)^{2n+2}{}_{\mu \nu}{}^{\rho \sigma }\, ,
\end{align}
\noindent
where we have exploited Eq. \eqref{eq:dualprop} to get rid of the Hodge star operators appropriately. Now, taking into account \eqref{recbn}, by virtue of \eqref{recursive1} we observe that \eqref{eq:eventerms} is indeed satisfied for $n+1$ as well. On the other hand, in order to see that \eqref{eq:oddterms} holds, it suffices to check that it satisfies the recursive relations \eqref{recursive1}. However, after noticing that
\begin{align}
\nonumber
\tensor{\left( \star \chi^{(1)} \right)}{_{\mu \nu}^{\alpha\beta}}\tensor{ \left( \star \chi^{(2n)} \right)}{_{\alpha \beta}^{\rho\sigma}}&= b_n \tensor{\left( \star \chi^{(1)} \right)}{_{\mu \nu}^{\alpha\beta}} \tensor{\left( \star \chi^{(1)} \right)}{_{\alpha \beta}^{\lambda \gamma}} \left(  \tensor{\chi}{^{(1)}}\right)^{2n-1}{}_{\lambda \gamma}{}^{\rho\sigma}\\&=-b_n  \tensor{\left( \star \chi^{(1)} \right)}{_{\mu \nu}^{\alpha\beta}}  \left(  \tensor{\chi}{^{(1)}}\right)^{2n-1}{}_{\alpha \beta}{}^{\lambda \gamma} \tensor{\left( \star \chi^{(1)} \right)}{_{\lambda \gamma}^{\rho \sigma}}=-\tensor{\left( \star \chi^{(2n)} \right)}{_{\mu \nu}^{\alpha\beta}}\tensor{ \left( \star \chi^{(1)} \right)}{_{\alpha \beta}^{\rho\sigma}}\,,
\end{align}
\end{widetext}
where have made a wide use of Eq. \eqref{eq:dualprop}, we realize that the recursive relations are identically satisfied. Hence \eqref{eq:eventerms} and \eqref{eq:oddterms} are the solution to the recursive relations \eqref{recursive1}. Since the $b_n$ are the coefficients of the Taylor series of $\sqrt{1+x^2}$, one can explicitly sum all tensors $\tensor{\chi}{^{(n)}_{\mu\nu}^{\rho\sigma}}$ to obtain:
\begin{equation}
\tensor{\chi}{_{\mu \nu}^{\rho \sigma}}=\alpha \tensor{\chi}{^{(1)}_{\mu \nu}^{\rho \sigma}}+\sqrt{\tensor{\delta}{_{[\mu}^{[\rho}} \tensor{\delta}{_{\nu]}^{\sigma]}}+\alpha^2 \left(  \tensor{\chi}{^{(1)}}\right)^{2}{}_{\mu \nu}{}^{\rho \sigma} }\,.
\label{eq:borninfeldgrav}
\end{equation}
By taking \eqref{eq:borninfeldgrav} into \eqref{eq:theory}, we find a fully non-perturbative and exactly duality-invariant theory with non-minimal couplings between gravity and electromagnetism. To the best of our knowledge, this is very first instance of such a theory and, interestingly enough, we observe that it possesses a high grade of resemblance with the usual (Einstein)Born-Infeld theories, as the Lagrangian involves the square root of certain quantity.  
We also find that, in this case, the susceptibility tensor $\tensor{\chi}{_{\mu\nu}^{\rho \sigma}}$ is only a function of the (traceless) Ricci tensor $\hat{R}_{\mu \nu}$. However, we note that replacing $\hat{R}_{\mu \nu}$ by any other symmetric and traceless tensor in \req{eq:chi1} would not alter the algebraic properties of $\tensor{\chi}{_{\mu\nu}^{\rho \sigma}}$, which would still respect duality. Therefore, we can generate any quadratic duality-invariant theory\footnote{Note that the parameter $\alpha$ can be reabsorbed in $\mathcal{T}_{\mu\nu}$.} by using \eqref{eq:borninfeldgrav} with
\begin{align}
\label{eq:chi12}
\tensor{\chi}{^{(1)}_{\mu\nu}^{\rho\sigma}}&=\tensor{\mathcal{T}}{_{[\mu}^{[\rho}}\tensor{\delta}{_{\nu]}^{\sigma]}}\, ,
\end{align}
where $\mathcal{T}_{\mu\nu}$ is an arbitrary symmetric and traceless tensor built out of the curvature. For the sake of concreteness, in the rest of the paper we focus on the theory generated by \eqref{eq:chi1}.

\textbf{Static and spherically symmetric configurations.}
Once we have obtained an exactly duality-invariant fully non-perturbative theory, our next objective will be to try to understand some features about its solutions. More concretely, we are going to focus on static and spherically symmetric (SSS) configurations, since they possess enough symmetry to be amenable to computations but still they are physically meaningful. They can in general be written in terms of the following ansatz

\begin{align}
\label{eq:sssansatz}
ds^2&=-N(r)^2f(r)dt^2+\frac{dr^2}{f(r)}+r^2\left(d\theta^2+\sin^2\theta d\phi^2\right)\, ,\\
F&=-A_t'(r)dt\wedge dr+p \sin\theta d\theta\wedge d\phi\, .
\label{eq:fsssansatz}
\end{align}
Here the metric depends on two functions $f(r)$ and $N(r)$, while $A_t(r)$ is the electrostatic potential and $p$ is a constant that represents the magnetic charge in Planck units. 

In order to compute the explicit form of the susceptibility tensor, note first that the traceless part of the Ricci tensor for an SSS metric reads
\begin{equation}
\tensor{\hat{R}}{^{\alpha}_{\beta}}=(X+Y) \tensor{\tau}{^{\alpha}_{\beta}}+(X-Y) \tensor{\rho}{^{\alpha}_{\beta}}-X\tensor{\sigma}{^{\alpha}_{\beta}}\, ,
\end{equation}
where 
\begin{eqnarray}
X&=&\frac{N(2f-2-r^2f'')-r(3r f' N'+2 f r N'')}{4r^2 N}\,  ,\quad \, \,\\
Y&=&-\frac{f N'}{rN}\, .
\end{eqnarray}
and where $\tau$, $\rho$ and $\sigma$ are the orthogonal projectors
\begin{equation}
\tensor{\tau}{^{\alpha}_{\beta}}=\tensor{\delta}{^{\alpha}_{t}}\tensor{\delta}{^{t}_{\beta}}\, ,\quad 
\tensor{\rho}{^{\alpha}_{\beta}}=\tensor{\delta}{^{\alpha}_{r}}\tensor{\delta}{^{r}_{\beta}}\, ,\quad 
\tensor{\sigma}{^{\alpha}_{\beta}}=\sum_{i=\theta,\phi}\tensor{\delta}{^{\alpha}_{i}}\tensor{\delta}{^{i}_{\beta}}\, .
\end{equation}
\noindent 
On the other hand, static and spherical symmetry force $\chi$ (and \emph{a fortiori} all the different $\chi^{(n)}$) to take the form
\begin{align}\notag
\tensor{\chi}{_{\mu\nu}^{\rho\sigma}}&=B \tensor{\tau}{_{[\mu}^{[\rho}}\tensor{\rho}{_{\nu]}^{\sigma]}}+C \tensor{\tau}{_{[\mu}^{[\rho}}\tensor{\sigma}{_{\nu]}^{\sigma]}}\\
&+D \tensor{\rho}{_{[\mu}^{[\rho}}\tensor{\sigma}{_{\nu]}^{\sigma]}}+E \tensor{\sigma}{_{[\mu}^{[\rho}}\tensor{\sigma}{_{\nu]}^{\sigma]}}\, ,
\label{eq:chican}
\end{align}
where $B$, $C$, $D$, $E$ are functions of $r$. Taking into account that 
\begin{align} \notag
\tensor{\delta}{_{[\mu}^{[\rho}}\tensor{\delta}{_{\nu]}^{\sigma]}}&=2  \tensor{\tau}{_{[\mu}^{[\rho}}\tensor{\rho}{_{\nu]}^{\sigma]}}+2 \tensor{\tau}{_{[\mu}^{[\rho}}\tensor{\sigma}{_{\nu]}^{\sigma]}}\\
&+2 \tensor{\rho}{_{[\mu}^{[\rho}}\tensor{\sigma}{_{\nu]}^{\sigma]}}+ \tensor{\sigma}{_{[\mu}^{[\rho}}\tensor{\sigma}{_{\nu]}^{\sigma]}}\, ,\\ \notag
\tensor{\chi}{^{(1)}_{\mu\nu}^{\rho\sigma}}&=-2 X \tensor{\tau}{_{[\mu}^{[\rho}}\tensor{\rho}{_{\nu]}^{\sigma]}}-Y \tensor{\tau}{_{[\mu}^{[\rho}}\tensor{\sigma}{_{\nu]}^{\sigma]}}\\
&+Y \tensor{\rho}{_{[\mu}^{[\rho}}\tensor{\sigma}{_{\nu]}^{\sigma]}}+X \tensor{\sigma}{_{[\mu}^{[\rho}}\tensor{\sigma}{_{\nu]}^{\sigma]}}\, .
\label{eq:chican1}
\end{align}
and that the projectors $\tau$, $\rho$ and $\sigma$ are mutually orthogonal, it is not difficult to obtain the coefficients $B$, $C$, $D$, $E$ from \eqref{eq:borninfeldgrav}. These take the following simple values:
\begin{eqnarray}
B&=&-2 \alpha X+2\sqrt{1+\alpha^2 X^2}\, , \\
C&=& -\alpha Y + 2 \sqrt{1+\frac{\alpha^2 Y^2}{4}}\,,\\
\label{eq:Esol}
E&=&\frac{2}{B}=\alpha X+\sqrt{1+\alpha^2 X^2}\,, \\
D&=&\frac{4}{C}=\alpha Y+ 2 \sqrt{1+\frac{\alpha^2 Y^2}{4}}\,,
\end{eqnarray}
Consequently, we have been able to find the exact form of the susceptibility tensor. 
This allows us to evaluate the reduced Lagrangian for the SSS ansatz given by \eqref{eq:sssansatz} and \eqref{eq:fsssansatz},  which takes the form
\begin{align}\notag
L&=\int d\theta d\phi\sqrt{|g|}\mathcal{L}\vert_{\mathrm{SSS}}\\
&=\frac{1}{4}\left[Nr^2R\vert_{\mathrm{SSS}}-2p^2\frac{NE}{r^2}+2(A_t')^2\frac{r^2}{NE}\right]\,.
\label{eq:lagsss}
\end{align}
Then we can find the equations of motion by varying this Lagrangian with respect to $A_{t}$, $f$ and $N$ \footnote{This process is indeed equivalent to computing first the complete Einstein and Maxwell equations and evaluating them on the SSS ansatz --- see Ref.~\cite{Cano:2020qhy} for a proof.}. The variation with respect to $A_{t}$ yields

\begin{equation}
\frac{\delta L}{\delta A_{t}}=-\frac{d}{dr}\left(\frac{A_{t}'r^2}{NE}\right)=0\, ,
\end{equation}
from where it follows that 

\begin{equation}
A_{t}'=\frac{ q N E}{r^2}\, ,
\end{equation}
where the integration constant $q$ represents the electric charge in Planck units.  On the other hand, taking the variation with respect to $f$ and $N$ and using the previous result, we find that the equations for the metric functions can be expressed as
\begin{eqnarray}
f-1+r f'&=&-(p^2+q^2)\frac{\delta}{\delta N}\left(\frac{NE}{r^2}\right)\, ,\\
r N'&=&(p^2+q^2)\frac{\delta}{\delta f}\left(\frac{NE}{r^2}\right)\, .
\end{eqnarray}
Therefore, they are manifestly invariant under a rotation of the charges $q$ and $p$, and it follows that the metric only depends on the combination $p^2+q^2\equiv \mathcal{Q}^2$. Due to the complicated form of $E$ in \req{eq:Esol}, these are highly-non linear fourth-order equations for $N$ and $f$, whose solution cannot be obtained analytically. However, for small $\alpha$ one can obtain the solution as a power series in this parameter. To order $\alpha^2$ it reads
\begin{eqnarray}
\notag 
f&=&1-\frac{2M}{r}+\frac{\mathcal{Q}^2}{r^2}-\frac{\left(7 \mathcal{Q}^4+5 \mathcal{Q}^2 r (2r-3M)\right) \alpha }{10 r^6}\\\notag &+&\frac{\left(5012 \mathcal{Q}^6+15 \mathcal{Q}^4 r
   (408 r-721 M)\right) \alpha ^2}{1680 r^{10}}+ \mathcal{O}(\alpha^3)\,, \\
   N&=&1+\frac{\mathcal{Q}^2 \alpha}{4 r^4}-\frac{41 \mathcal{Q}^4 \alpha^2}{32 r^8}+ \mathcal{O}(\alpha^3)\, ,
   \label{eq:pertsol}
\end{eqnarray}
where $M$ is the mass. We can see this solution is a deformation of the Reissner-Nordstr\"om one. However, the perturbative expansion in $\alpha$ is only valid as long as the corrections are small and hence we cannot see what happens to black holes when $\alpha\sim \mathcal{Q}^2$. In that regime, one would need to resort to numeric methods to solve the equations of motion. 

\textbf{Extremal black holes and near-horizon geometries.}
Fortunately, the situation improves if we are interested in extremal black holes. In that case, it is possible to obtain the near-horizon metric as well as the black hole entropy by using Sen's method \cite{Sen:2005iz,Sen:2007qy}. This method essentially consists in evaluating the Lagrangian on an AdS$_2\times\mathbb{S}^2$ geometry. The near-horizon solution is then obtained by extremizing the action, while the entropy is given by the Legendre transform of the Lagrangian with respect to the electric field. We follow this process in detail next. 

%This is an algorithm by which the near-horizon geometry of extremal black holes as well as its entropy are obtained by extremizing a single function that is called entropy function. More concretely, the metric coefficients and the electromagnetic charges are computed through this optimization procedure, while the entropy is calculated by evaluating the entropy function on the previously worked out extremum. %On implementing Sen's method for the case at hands, we will try to be as illustrative as possible, but in any case we would like to refer the reader to \cite{Sen:2007qy}, where a self-contained and complete exposition of such method is presented.

We start by considering the following AdS$_2\times\mathbb{S}^2$ ansatz for the metric and the field strength:
\begin{align}
\label{eq:exnhansatz}
ds^2&=a\left (-\rho^2 d t^2+\frac{1}{\rho^2}d \rho^2 \right) + b\left ( d \theta^2+\sin^2 \theta d \varphi^2 \right ) \, , \\
F&=-e d t \wedge d \rho+p \sin \theta d \theta \wedge d \varphi\,.
\end{align}
Here $a=R_{\rm AdS_{2}}^2$, $b=R_{\mathbb{S}^2}^2$ are the radii squared of the AdS$_2$ factor and of the black hole horizon, respectively, $p$ is the magnetic charge and $e$ will be related to the electric charge. This geometry can be obtained from the general SSS ansatz in Eqs.~\req{eq:sssansatz} and \req{eq:fsssansatz} by setting $r=\sqrt{b}+\rho$, $f=\rho^2/a$, $N=a$, $A_t'=e$ and keeping the leading terms in the expansion around $\rho\rightarrow 0$. Thus, the reduced Lagrangian reads in this case
\begin{equation}
L(a,b,e,p)=\frac{1}{2} \left(a-b-p^2\hat E+e^2\frac{1}{\hat E}\right)\,, 
\end{equation}
where 

\begin{equation}
\hat E=-\alpha \frac{a+b}{2b^2}+ \frac{a}{b}\sqrt{1+\alpha^2 \frac{(a+b)^2}{4 a^2 b^2}}\, .
\end{equation}
The entropy function $\mathcal{E}(a,b,e;q,p)$ is then defined as:
\begin{equation}
\mathcal{E}(a,b,e;q,p)=2 \pi(e q-L(a,b,e,p))\,,
\label{eq:entrfunc}
\end{equation}
where $q$ is the electric charge of the configuration. 
Extremizing the entropy function with respect to $a$, $b$ and $e$ yields the equations satisfied by $a$ and $b$ as well as the relation between the electric charge and $e$.
Indeed, the equation $\partial\mathcal{E}/\partial e=0$ yields
\begin{equation}
e=\hat E q\, .
\label{eq:seneq1}
\end{equation}
On the other hand, by deriving with respect to $a$ and $b$ and using this result we obtain the following sets of equations for $a$ and $b$:
\begin{eqnarray}
\label{eq:seneq2}
\frac{1}{\pi}\frac{\partial \mathcal{E}}{\partial a}&= &-1+(p^2+q^2)\frac{\partial \hat{E}}{\partial a}=0\, ,\\
\label{eq:seneq3}
\frac{1}{\pi}\frac{\partial \mathcal{E}}{\partial b}&= &1+(p^2+q^2)\frac{\partial \hat{E}}{\partial b}=0\, .
\end{eqnarray}
We observe again that the equations are invariant under a rotation of the electric and magnetic charges. Notice also that these equations are highly nonlinear --- in fact, they are not even polynomial --- due to the form of $\hat E$ given above. In spite of this, these equations can be solved in full generality and we observe that they admit four different solutions. However, there is only one solution with $a,b>0$, and it is given by
\begin{eqnarray}
\label{eq:asen}
a&=&\frac{1}{2}\left ( p^2+q^2+\alpha+ \sqrt{(p^2+q^2)^2-\alpha^2}\right )\, ,\\
 b&=&\frac{1}{2}\left ( p^2+q^2- \alpha+ \sqrt{(p^2+q^2)^2-\alpha^2}\right )\,.
\label{eq:bsen}
\end{eqnarray}
Interestingly, this implies that $\hat E=1$ and hence $e=q$. 
Finally, substituting these values for $a,b,e$ in the entropy function \eqref{eq:entrfunc} we arrive to the following result for the entropy of these extremal black holes:
\begin{equation}
S=\pi(p^2+q^2-\alpha)\,.
\end{equation}
Surprisingly enough, we find that there is only a constant correction to the entropy with respect to the Einstein-Maxwell value --- we remark that this is the exact value of the entropy and not just an approximation. Notice that, even when one adds only a finite number of higher-order terms in the action, the entropy (and the rest of the quantities) will be typically modified by an infinite tower of $\alpha$ terms. Here we observe the opposite: the action contains an infinite number of higher-order terms as dictated by duality invariance, but in turn the entropy only has a correction of order $\alpha$. 

Let us take a closer look at this solution. While the entropy is finite and real for any value of the charges, we see that this is not the case for $a$ and $b$. In fact, for any sign of $\alpha$ we see that these extremal geometries only exist for
\begin{equation}
p^2+q^2\ge |\alpha|\, . 
\end{equation} 
Therefore, there is a minimum amount of charge needed to produce an extremal black hole, implying that all black holes with $p^2+q^2<|\alpha|$ must be necessarily non-extremal. On the other hand, the properties of these black holes near the minimal charge are quite different depending on the sign of $\alpha$. When $\alpha>0$, the radius of the AdS$_2$ tends to the constant value $a=\alpha$ as $p^2+q^2\rightarrow \alpha$, while the area of the horizon and the entropy vanish in this limit. In the case of $\alpha<0$ we observe the contrary: the radius of AdS$_2$ goes to zero, while both the entropy and the area tend to a constant value, namely, $S=A/2=2\pi \vert \alpha\vert $, as $p^2+q^2\rightarrow \vert \alpha \vert$.

In order to determine the sign of $\alpha$, one may use the so-called mild form \cite{Cheung:2018cwt,Hamada:2018dde} of the weak gravity conjecture \cite{ArkaniHamed:2006dz}. This states that the corrections to the mass of extremal black holes must be non-positive, so that the decay of an extremal black hole into a set of smaller black holes is possible. The near-horizon geometry does not allow one to obtain the mass of the black hole, but we can obtain it from the perturbative solution \req{eq:pertsol}. Imposing the extremality condition $f(r_+)=f'(r_+)=0$, we find that 
\begin{eqnarray}
\label{eq:extmass}
M_{\mathrm{ext}}&=&\mathcal{Q}-\frac{\alpha}{10 \mathcal{Q}}-\frac{\alpha^2 }{84 \mathcal{Q}^3}+\mathcal{O}(\alpha^3)\, ,
\end{eqnarray}
while the extremal radius $r_{+}$ agrees exactly with the expansion of $\sqrt{b}$ in \req{eq:bsen}. We have checked that the $\alpha$-expansion of the mass converges very rapidly and the expression above turns out to be very accurate even for $\mathcal{Q}=\sqrt{|\alpha|}$. We see that in order for the corrections to the mass to be non-positive we must have $\alpha\ge0$. Then, at the minimal charge $\mathcal{Q}^{\rm min}=\sqrt{\alpha}$ the mass becomes $M_{\mathrm{ext}}^{\rm min}\approx 0.88 \sqrt{\alpha}$ and the entropy and area of extremal black holes vanish.

\textbf{Discussion.}
Duality-invariant modifications of Einstein-Maxwell theory are interesting and well-motivated theories, but they are highly nonlinear and often one cannot perform exact computations. In this letter we have provided the first example of exactly-duality invariant theories with non-minimal couplings. Namely, we have shown that, in the case of a quadratic dependence on the field strength, these theories have the form given by  \eqref{eq:theory}, \eqref{eq:borninfeldgrav} and \eqref{eq:chi12}. It would also be interesting to look for more general nonminimal duality-preserving theories, \textit{i.e.}, including as well higher powers of the Maxwell field strength, but this is a challenging problem which will be treated elsewhere.
 
%In this letter we have shown that one can make sense of these theories even at a non-perturbative level by providing an exact computation of black hole solutions in a nonminimal theory with exact duality invariance. Our results open up the road for computing black hole solutions in more general models. In particular, the method employed to sum the whole series of higher-derivative terms can be applied in full generality to any other theory of the form \req{eq:theory}.  Another obvious extension of our analysis entails considering asymptotically AdS or dS solutions by including a cosmological constant in the action. 

Focusing on the simplest of these theories, we have studied its static and spherically symmetric solutions. 
As we have shown, the equations of motion satisfied by the metric in the latter theory are invariant under rotations of the electric and magnetic charges, but due to their complexity they can only be solved analytically in the perturbative regime --- see  \req{eq:pertsol}. 
However, we found that the near-horizon geometry of extremal black holes can be obtained exactly. 
%\req{eq:theory}

A remarkable aspect about these extremal black holes is that their entropy only receives a constant correction, which is striking since the action is modified in a very nonlinear way. A similar result is observed in the case of Einstein-Born-Infeld theory, which suggests that duality somehow simplifies the corrections to the entropy. It would be interesting to explore other theories to understand this possible connection better, but we do not have as of this moment a simple explanation for this observation.

In addition, these extremal black holes possess a minimal charge below which no solutions exist. Thus, it would follow that any black hole with a charge below this minimum value must be non-extremal --- no matter how small the mass is.  
We have also shown that the weak gravity conjecture imposes the coupling constant $\alpha$ to be positive, which led us to the conclusion that, at the minimal charge, the area and entropy of extremal black hole vanish. This is an intriguing behavior, and it is tantalizing to assume that this minimal charge coincides precisely with the elementary electric charge. An extremal black hole with the charge of an electron is trivially the one with the lowest (non-zero) charge, and one could argue that its entropy would vanish because it would contain only one microstate. However, we note that the entropy can always be shifted by the introduction of a topological Gauss-Bonnet term in the action, so the entropy of the minimal extremal black hole can be changed. 

These issues could be better understood by trying to embed this theory in string theory, in whose case, a precise entropy counting is available, \textit{e.g.} \cite{Strominger:1996sh,Kutasov:1998zh,Dabholkar_2005,Sen:2007qy, Prester_2008}. In fact, we have checked that our solution \req{eq:pertsol} coincides with the $\alpha'$-corrected Reissner-Nordstr\"om black hole of Ref.~\cite{Cano:2019ycn}, upon the identification $\alpha=\alpha'/8$ \footnote{More precisely, both metrics are related by a redefinition $g_{\mu\nu}^{\rm ST}=g_{\mu\nu}^{\rm ours}+3\alpha T_{\mu\nu}$ to first order in $\alpha$, but this only means that the theories are written in different frames.}. This shows that our theory \req{eq:theory} captures some of the stringy $\alpha'$-corrections, at least in the situations where the additional degrees of freedom besides the metric and the electromagnetic field can be neglected.

\vspace{0.4cm}
\begin{acknowledgments}   
\textbf{\textit{Acknowledgments.}} PAC would like to thank M. D. S\'anchez for her help in reviewing this paper. 
The work of PAC is supported by a postdoctoral fellowship from the Research Foundation - Flanders (FWO grant 12ZH121N). The work of \'AM is funded by the Spanish FPU Grant No.  FPU17/04964 and by the Deutscher Akademischer Austauschdienst (DAAD), through the Short-Term Research Grant No. 91791300. \'AM was further supported by the MCIU/AEI/FEDER UE grant PGC2018-095205-B-I00 and by the ``Centro de Excelencia Severo Ochoa'' Program grant SEV-2016-0597.

\end{acknowledgments}
\bibliographystyle{apsrev4-1} % Tell bibtex which bibliography style to use
%\vspace{1cm}
\bibliography{Biblio}

\end{document}
%
% ****** End of file template.aps ******